\documentclass[amssymb,prb,twocolumn,showpacs,cite]{revtex4}
\usepackage{amsmath}
\usepackage{graphicx}
\usepackage{verbatim}
\usepackage{graphics}
\begin{document}

\title{
Control of Spin Blockade by ac Magnetic Fields in Triple Quantum Dots
}
\author{Maria Busl$^1$, Rafael S\'anchez$^2$ and Gloria Platero$^1$}
\affiliation{$^1$Instituto de Ciencia de Materiales de Madrid, CSIC, Cantoblanco, 28049, Madrid,
Spain\\
$^2$ D\'epartement de Physique Th\'eorique, Universit\'e de Gen\`eve,
CH-1211 Gen\`eve 4, Switzerland}
\begin{abstract}
\vspace{0.5cm} We analyze coherent spin phenomena in triple quantum
dots in triangular configuration under crossed dc and ac magnetic
fields.
In particular, we discuss the interplay between Aharonov-Bohm
current oscillations, coherent electron trapping and spin blockade
under two-electron spin resonance configurations. We demonstrate
an unexpected antiresonant behaviour in the current, allowing for
both removal and restoration of maximally entangled spin blockaded
states by tuning the ac field frequency.
Our theoretical predictions indicate how to
manipulate spin qubits in a triangular quantum dot array.
\end{abstract}
\maketitle

Electronic transport through mesoscopic systems can become
correlated not only by charge interaction but also by the spin
degree of freedom. A dramatic combination of both can be found in
systems where strong Coulomb interaction limits the population to a
small number of electrons (Coulomb blockade) and where Pauli
exclusion principle avoids certain internal transitions --- spin
blockade (SB). This was first observed as a rectification effect in the
current through a double quantum dot (DQD)\cite{ono}.\\
Recent experiments have taken advantage of SB to achieve qubit operations in a double dot by electric gate control\cite{petta} or by electron spin resonance (ESR)\cite{koppens}.
It consists in inducing transitions between the electron's spin-up
and spin-down states, which are splitted by the Zeeman energy coming
from a dc magnetic field, $B_\text{dc}$.
Different mechanisms have been considered: crossed dc and ac
magnetic fields ($B_\text{ac}$), where the ac frequency is resonant
with the Zeeman splitting\cite{loss1, rafa}, effective $B_\text{ac}$
induced by ac electric fields in the presence of spin-orbit
interaction\cite{nowack}, slanting Zeeman fields\cite{pioroladriere} or hyperfine interaction\cite{laird}.

Lately, a next step towards quantum dot arrays has been reached: tunnel spectroscopy
measurements with triple quantum dots (TQD), both in series\cite{ludwig}
and in triangular configurations\cite{haug} have been achieved.
Theoretical works in these systems\cite{hawrylak,bulka} analyze
their eigenstates
and stability diagram, as well as the
effect of a magnetic field
penetrating the structure. TQDs with strongly correlated
electrons have also been
investigated in the Kondo regime\cite{kikoin} and have been proposed as spin
entanglers\cite{saraga}. Additionally, these systems show a more peculiar property which is intrinsic to three-level systems, namely
coherent population trapping, which is  a well-known effect in quantum optics and which was observed in three-level
atoms excited by two resonant laser fields\cite{optics}. There, the
electronic wave function evolves towards an eigenstate superposition,
a so-called {\it dark state}, which is decoupled from the laser fields and therefore it manifests as an antiresonance in the emission spectrum.
An analogy in transport has been made when coherent superpositions avoid transport by interference between tunneling events.
These dark states can be achieved by driving three-level double dots with bichromatic ac electric fields\cite{brandes} or by the interference of tunneling processes in TQDs\cite{emary,wir}. It was shown\cite{emaryAB} how coherent trapping can be lifted in closed-loop TQDs by means of the Aharonov-Bohm (A-B) effect\cite{bohm}.

\begin{figure}[t]
\begin{center}
\includegraphics[width=\linewidth] {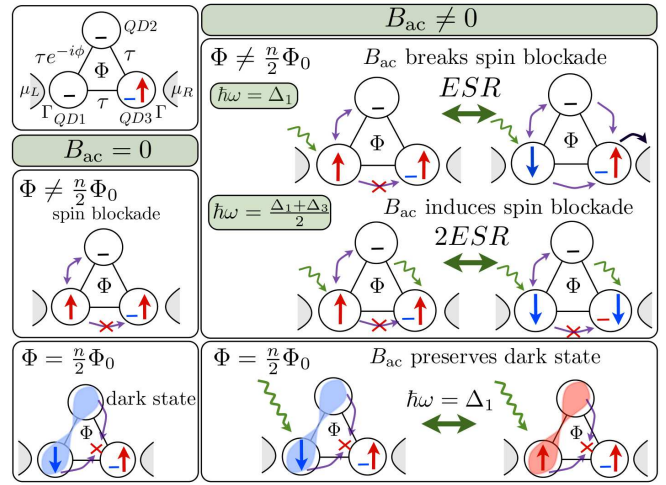}
\end{center}
\caption{\label{fig1}\small
(Color online) Coherent processes in a triangular TQD, 
with one electron confined in the dot connected to the drain.
$\Delta_1 = \Delta_2\neq \Delta_3$, where $\Delta_i$ is the Zeeman splitting in
dot $i$.
Transport through the system depends on the magnetic flux $\Phi$ 
penetrating the system
and on the frequency $\omega$ of the time-dependent
magnetic field $B_{\text{ac}}$. The shaded regions indicate the existence of a dark state.}
\end{figure}

Here, we will discuss the electron
spin dynamics and transport for the case where a triangular  TQD
 contains up to two extra electrons, as shown in Fig.~\ref{fig1}. In contrast to the
single electron case, spin correlations can influence transport due to SB\cite{spincorr}.
We will show that at certain $B_{\text{ac}}$ frequencies and sample configurations, the magnetic field brings the electronic wave function into a superposition of parallel spins states,
 unexpectedly bringing the system back to SB.

{\it Model} --- We  consider a system consisting of three dots which are
coupled through tunnel barriers,
and dots 1 and 3 are also connected to source and drain contacts
respectively. The  Hamiltonian of the system is
$\hat H(t) = \hat H_{\text{TQD}}+\hat H_{\tau}+\hat H_{\text{T}}+\hat H_{\text{leads}}+\hat H_{\text{B}}(t).$
$\hat H_{\text{TQD}}=\sum_{i\sigma}\epsilon_{i}\hat{c}^{\dagger}_{i\sigma}
\hat{c}_{i\sigma}+\sum_{i}U_{i}\hat{n}_{i}\hat{n}_{i}
+\sum_{i,j, i\ne j}V_{ij}\hat{n}_{i}\hat{n}_{j}$ describes the uncoupled TQD,
$\hat H_{\tau}=-\sum_{ij\sigma}(\tau_{ij}\hat{c}^{\dagger}_{i\sigma}
\hat{c}_{j\sigma}+\text{h.c.})$, the coherent tunneling between the dots,
$\hat H_{\text{T}}=\sum_{l\in{L,R}k\sigma}(\gamma_{l}\hat{d}^{\dagger}_{lk\sigma}
\hat{c}_{l\sigma}+\text{h.c.})$ describes the
coupling of the dots to the leads,
and $\hat H_{\text{leads}}=\sum_{lk\sigma}\epsilon_{lk}\hat{d}^{\dagger}_{lk\sigma}
\hat{d}_{lk\sigma}$ the leads
themselves. $\epsilon_{i}$ is the energy of an electron
located in dot $i$, $U_i$ is the intra-dot and $V_{ij} = V$ the
inter-dot Coulomb repulsion. 
If not stated otherwise, we set $|\tau_{ij}| = \tau$.

The Hamiltonian for the magnetic field, $\hat H_{\text{B}}(t)$, has two components:
a time-independent dc component along the $z$-axis that breaks the spin-degeneracy by a
Zeeman-splitting $\Delta_{i}=g_iB_{zi}$, and a circularly polarized
ac component in the $xy$-plane that rotates the $z$-component of
the electron spin when its frequency fulfills the resonance condition
$\hbar\omega_i=\Delta_i$. It reads 
$\hat H_{\text{B}}(t)=\sum_{i=1}^3[\Delta_i\hat S_{zi}+B_{\text{ac}}(\cos(\omega t)\hat S_{xi}+
\sin(\omega t)\hat S_{yi})]$,
being
${\bf S}_{i}=\frac{1}{2}\sum_{\sigma\sigma'}\hat{c}^{\dagger}_{i\sigma}
\sigma_{\sigma\sigma'}\hat{c}_{i\sigma'}$
the spin operator of the $-i$th dot. As shown in \cite{rafa}, the ac magnetic field has no effect on SB unless the Zeeman splitting is inhomogeneous in the sample. Here we will consider the simplest configuration that allows to analyze the relevant mechanisms: $\Delta_1=\Delta_2\ne\Delta_3$.
The experimental feasibility\cite{huang,pioroladriere} justifies the choice of the present configuration.

The dynamics of the system is given by the time evolution of the
reduced density matrix whose equations of motion read as, within the
Born-Markov-approximation:
\begin{eqnarray}
\dot{\rho}_{ln}(t)=&-i&\langle l|[\hat H_{\text{TQD}}+\hat H_{\tau}+\hat H_{\text{B}}(t),
\rho]|n\rangle\\
&+&\sum_{k\neq{n}}(\Gamma_{nk}\rho_{kk}-\Gamma_{kn}\rho_{nn})\delta_{ln}-\Lambda_{ln}\rho_{ln}(1-\delta_{ln})\nonumber
\label{density}
\end{eqnarray}
The commutator accounts for the coherent dynamics in the TQD,
$\Gamma_{ln}$ are the transition rates from state $|n\rangle$ to
state $|l\rangle$ induced by the coupling to the
leads and decoherence appears by
$\Lambda_{ln}=\frac{1}{2}\sum_{k} (\Gamma_{kl}+\Gamma_{kn})$.

We consider a configuration where the dot coupled to the drain 
is permanently occupied by one electron (see Fig.~\ref{fig1}),
and only up to two electrons can be in the system. 
Double occupancy is only allowed in the drain dot.
This is the case 
when the chemical potentials in the leads satisfy 
$\epsilon_3+V<\mu_R<\epsilon_3+U_3$ and $\mu_L<\epsilon_1+2V$.
For resonant tunneling, $\epsilon_1 = \epsilon_2$
and $\epsilon_{1,2} + V = \epsilon_{3} + U_{3}$.  
Out of the full TQD basis with up to two electrons, there are then 
eleven one- and two-electron states that dominate the dynamics: 
$|0,0,\sigma\rangle$,
$|\chi_{1\sigma\sigma'}\rangle=|\sigma,0,\sigma'\rangle$, $|\chi_{2\sigma\sigma'}\rangle=|0,\sigma,\sigma'\rangle$ and $|S_3\rangle=|0,0,{\uparrow}{\downarrow}\rangle$,
with $\sigma,\sigma'=\{{\uparrow},{\downarrow}\}$.
Transport is biased from left to right and only state $|S_3\rangle$ contributes to 
tunneling through to the drain, acting as a bottleneck for the current:
$I(t)=\sum_n\Gamma_{nS_3}\rho_{S_3S_3}(t)$.
Though being confined, the electron in dot 3
is essential to induce spin correlated transport.
A $B_{\text{dc}}$ perpendicular to the plane of the triangular dot
structure (Fig.~\ref{fig1}) encloses a magnetic flux $\Phi$ such
that electron tunneling acquires an additional phase
$\phi=2\pi\Phi/\Phi_0$, with $\Phi_0=h/e$ being the flux quantum. 
We accumulate the phase between dot 1 and dot 2, 
$\tau_{12}=\tau e^{{-{i}\phi}}$.

\begin{figure}[t]
\begin{center}
\includegraphics[width=\linewidth,clip]{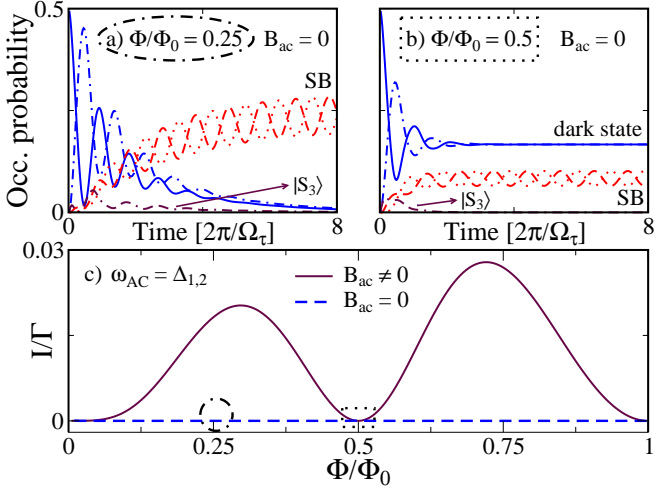}
\end{center}
\caption{\label{fig2}\small (Color online) 
(a) $\rho_{ii}$ for $\Phi/\Phi_0 =
0.25$, $B_{\text{ac}}=0$.
$I$ is blocked due to SB, once the parallel spin states
(dashed and dotted red lines)
are occupied.
(b) $\rho_{ii}$ for $\Phi/\Phi_0 = 0.5$,
$B_{\text{ac}}=0$. 
Due to SB, there is a
finite occupation of parallel spin states $|\chi_{l\sigma\sigma}\rangle$ (dashed and dotted red lines),
while electrons with antiparallel spin form dark states as in \eqref{ds} (solid and dashed-dotted blue lines), all of them contributing to quench the current.
(c) $I$-$\Phi$: for $B_{\text{ac}}=0$, $I=0$ due to SB (dashed blue line);
$B_{\text{ac}}\neq 0$: for $\hbar\omega_{\text{ac}}=\Delta_{1,2}$, SB is removed
and the current shows A-B-like oscillations (solid purple line).
Rabi frequency: $\Omega_{\tau} = 2\tau$. $\tau=0.0026$,
$\Gamma=0.01$ in meV, $\Delta_3=0.77 \Delta_1$, $\Delta_1=\Delta_2$.}
\end{figure}

\emph{Undriven case ($B_{\text{ac}} = 0$):} --- It is well
known\cite{emaryAB} for a TQD with up to one extra electron, that
due to interference, the current oscillates with $\Phi$
(A-B oscillations) and periodically drops to zero with a periodicity
of $\Phi_0/2$. 
For the understanding of the two-electron spin dynamics, it is crucial to look at
the eigenstates of this system, which change depending on the flux $\Phi$. 
For $\Phi/\Phi_0 = \frac{n}{2}$
\begin{align}
|\psi_{\sigma\sigma'}^{-}\rangle &=
\frac{1}{\sqrt{2}}(|\chi_{2\sigma\sigma'}\rangle-|\chi_{1\sigma\sigma'}\rangle) &\sigma, \sigma' = \{\uparrow,\downarrow\} \label{ds} \\
|\psi_{\sigma\sigma}^{+}\rangle &=
\frac{1}{\sqrt{2}}(|\chi_{2\sigma\sigma}\rangle+|\chi_{1\sigma\sigma}\rangle) &\sigma = \{\uparrow,\downarrow\} \label{ds2}
\end{align}
are eigenstates of the closed system. States (\ref{ds}) avoid 
tunneling to $|S_3\rangle$:
$\langle\psi_{\sigma\sigma'}^{-}|\hat H_{\tau}|S_3\rangle=0$, which is why they are also called \emph{dark states},
see Fig.~\ref{fig1}. Occupation of $|S_3\rangle$ thus decays by the coupling to the drain (Fig.~\ref{fig2}b) 
and current is blocked. The states in \eqref{ds} remind of the dark states found in the single electron case\cite{emary}.
A significant difference is that \emph{for two electrons the spin degree of freedom plays a role}:
Pauli exclusion principle introduces spin correlation such that dark states $|\psi_{\sigma\sigma'}^{-}\rangle$ with $\sigma = \sigma'$
are avoided. The electrons are rather being trapped in combinations of dark states $|\psi_{\sigma\sigma'}^{-}\rangle$ 
with $\sigma\neq\sigma'$ and
spin-blockaded states $|\psi_{\sigma\sigma}^{+}\rangle$.
Thus, SB competes with coherent population trapping in the blocking of the current, and the relative occupation of $|\psi_{\sigma\sigma'}^{-}\rangle$ ($\sigma\neq\sigma'$), and $|\psi_{\sigma\sigma}^{+}\rangle$ depends on the initial condition.
If however $\Phi=\Phi_0 /4$, the A-B phase removes the dark state 
and only eigenstates with parallel spins are decoupled from $|S_3\rangle$:
\begin{equation}
|\xi_{\sigma\sigma}^{\pm}\rangle=
\frac{1}{\sqrt{2}}(|\chi_{2\sigma\sigma}\rangle\pm\dot{\imath}|\chi_{1\sigma\sigma}\rangle) \  \  \  \ \sigma = \{\uparrow,\downarrow\} 
\label{sb}
\end{equation}
Coherent trapping is hence lifted, however, transport is still
cancelled by SB (Figs.~\ref{fig1}, \ref{fig2}a).
One can appreciate that without $B_{\text{ac}}$, the system is
always blocked for transport --- the stationary current is
insensitive to A-B effect due to SB.

\emph{Driven case ($B_{\text{ac}}\ne 0$):} --- In order to remove
SB, we apply a time-dependent $B_{\text{ac}}$. 
Fig.~\ref{fig2}c shows the $I$-$\Phi$ characteristics of the TQD
excited by $B_{\text{ac}}$, where for every value of
$\Phi$, the magnetic field frequency fulfills the resonance
condition $\hbar\omega = \Delta_{1,2}$.
 For $\Phi/\Phi_0\ne n/2$, dark states are avoided by A-B effect, and
$B_{\text{ac}}$ enables transitions of the form
$|\chi_{l\sigma\sigma}\rangle\rightarrow|\chi_{l\sigma'\sigma}\rangle\rightarrow|S_3\rangle$
that produce a finite current.\\
It can be shown that
$B_{\text{ac}}$ does not affect the destructive tunneling interference of the superpositions \eqref{ds}.
Then, if $\Phi/\Phi_0=n/2$ the system evolves towards a state which is only composed of dark states performing spin rotations: 
$|\psi_{\sigma\sigma'}^{-}\rangle\leftrightarrow|\psi_{\sigma\sigma}^{-}\rangle$, as shown schematically in Fig.~\ref{fig1}. 
Since the dark states are 
decoupled from transport, the oscillations can only be affected by decoherence 
due to spin scattering processes, which are not considered here.
Hence, a $B_{\text{ac}}$ induces current through the system only when assisted by the A-B
lifting of dark states --- i.e. for $\Phi/\Phi_0\ne n/2$ (Fig.~\ref{fig2}c).

\begin{figure}[t]
\begin{center}
\includegraphics[width=\linewidth]{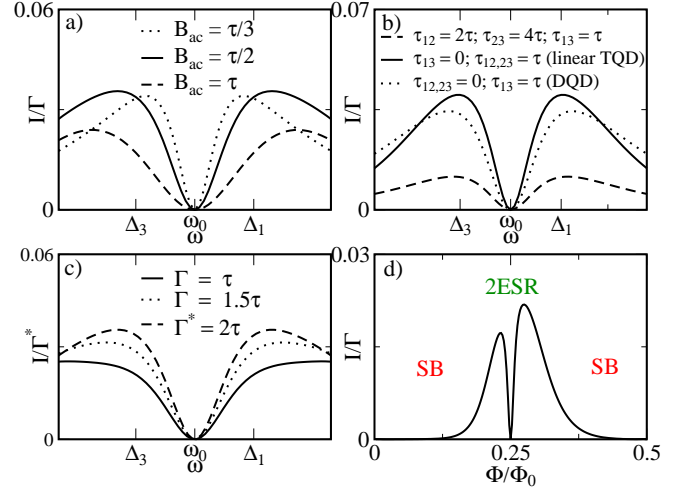}
\end{center}
\caption{\label{fig3}\small
2ESR in a TQD by tuning $\omega$, for $\Phi/\Phi_0 = 0.25$,
 is manifested in the current as an antiresonance at 
 $\hbar\omega_0=(\Delta_1+ \Delta_3)/2$.
(a) For different $B_\text{ac}$ and fixed $\tau_{ij}=\tau$: The width of the antiresonance depends on the
Rabi frequencies associated with $\tau$ and $B_{\text{ac}}$. (b) For
different $\tau_{ij}$ and fixed $B_\text{ac}=\tau/2$. The antiresonance appears for
any configuration of the $\tau_{ij}$.
(c) For different $\Gamma$ and $\tau_{ij}=\tau$.
(d) $I$-$\Phi$ for fixed $\omega$ while tuning $B_{\text{dc}}$, so both
$\Phi$ and $\Delta_i$ are modified.  A-B oscillations are suppressed
by SB except when the Zeeman splittings are close to
resonance with $\omega$. At $\Phi/\Phi_0 = 0.25$, 
$\omega = \omega_0$ and current vanishes. 
Parameters: $\tau = 0.005$, $\Gamma = \Gamma^{*} = 0.01$.}
\end{figure}

Remarkably,
not imposing the resonance condition $\hbar\omega=\Delta_i$,
one can find a novel kind of SB induced by $B_{\text{ac}}$, quenching the current even
in the presence of A-B effect.
This is the main result of our work.
As can be seen in Fig.~\ref{fig3}a, the current shows a resonant
behaviour as the frequency of $B_\text{ac}$ approaches the ESR
condition (i.e. $\hbar\omega\sim\Delta_1, \Delta_3$). Surprisingly though,
an antiresonance appears for $\hbar\omega_0=(\Delta_1+ \Delta_3)/2$, i.e. when the
two electrons are equally far from the ESR condition. Note
that the two peaks around the antiresonance are not Lorentzian-like
and cannot be identified as two different resonance peaks centered at the
conditions $\hbar\omega=\Delta_1=\Delta_2$ and
$\hbar\omega=\Delta_3$, but as a collective effect due to the
simultaneous rotation of the {\it two} electron spins (2ESR), cf. Fig~\ref{fig1}.\\
We want to stress that the {\it appearance} of the antiresonance does
not depend on the field intensity $B_{\text{ac}}$ or tunnel couplings $\tau_{ij}$
(see Fig.~\ref{fig3}a,b): it occurs for different $\tau_{ij}$ as well as for
linear TQD configurations (setting $\tau_{13}=0$) and DQDs in series\cite{rafa} 
(setting $\tau_{12}=\tau_{23}=0$), see Fig.~\ref{fig3}b.
The
{\it width} of the antiresonance scales with the Rabi frequency of
the coherent processes involved\cite{tobias}: spin
rotation ($\propto B_{\text{ac}}$) and interdot tunneling
($\propto \tau_{ij}$) (Figs.~\ref{fig3}a and
\ref{fig3}b, respectively); it also depends on the tunneling rate through
the contact barriers, which induce decoherence, see Fig.~\ref{fig3}c.

The quenching of the current can be understood analytically by transforming the
Hamiltonian into the rotating frame. Applying the unitary operator
$\hat U(t)=\text{exp}\{-i\omega t \sum_{i=1}^3\hat S_{zi}\}$,
the magnetic field term reads:
$\hat H_\text{B}'=\sum_{i=1}^3[(\Delta_i-\hbar\omega)\hat S_{zi}+B_{\text{ac}}\hat S_{xi}].$
One can easily
verify that, at $\hbar\omega_0=(\Delta_1+ \Delta_3)/2$,
the coherent superpositions
\begin{equation}
|\Psi^{\pm}\rangle = \frac{1}{\sqrt{2}}\left(|\xi_{\downarrow\downarrow}^{\pm}\rangle - |\xi_{\uparrow\uparrow}^{\pm}\rangle\right)
\label{bacstate}
\end{equation}
are eigenstates of the Hamiltonian $\hat H' = \hat H_{\tau} + \hat H'_{\text{B}}$.
Since the electrons in (\ref{bacstate}) have parallel spins, 
current is quenched due to SB.
Note that the electron spins in (\ref{bacstate}) are maximally
entangled. We want to emphasize that SB can be {\it
switched} on and off by tuning the frequency of $B_{\text{ac}}$, which is usually introduced to lift it,
or by changing the flux $\Phi$ at a fixed frequency $\omega$, see Fig.~\ref{fig3}d. 

In TQDs, a necessary condition for (\ref{bacstate})
to be eigenstates of $\hat H'$ and thus for the current blocking to occur, 
is the equal coupling of dots 1 and 2 to
$B_{\text{dc}}$, i. e. $\Delta_1=\Delta_2$ $(\neq\Delta_3)$.
If $\Delta_1\ne\Delta_2$ though, this symmetry is broken and
$\hat H'_{\text{B}}$ couples all parallel to antiparallel spin states and 
thus to the transport state $|S_3\rangle$.
However, numerical results show that even in the asymmetric case, 
a pronounced antiresonance due to SB still appears in the current.
By means of a perturbative analysis for $\Delta_1-\Delta_2 \ll \tau$ it can be shown that 
the antiresonance occurs at a frequency $\omega_1 \approx \frac{1}{2}(\frac{\Delta_1 + \Delta_2}{2} + \Delta_3)$,
see Fig.~\ref{fig4}.
The electrons drop into an eigenstate $|\Psi^{\star}\rangle$ which is similar to (\ref{bacstate}) but includes a 
small contribution of antiparallel spin states which produces a small leakage current. 
This leakage current increases as  $\Delta_1 - \Delta_2$ becomes of the order of $\tau$.

{\it Bichromatic $B_{\text{ac}}$} --- Finally we will show that for
$\Delta_1 = \Delta_2$
SB can also be induced by a bichromatic $B_{\text{ac}}$, tuning its frequencies to $\hbar\omega_{1}=\Delta_{1,2}$ and
$\hbar\omega_{2}=\Delta_{3}$, so every electron is kept in resonance regardless of its location. Assuming that the inhomogeneity in the
Zeeman splittings is high enough so one can neglect the
off-resonance terms, the Hamiltonian $\hat H_{\text{B}}(t)$ can be written as
$\hat H_{\text{B,2}}=\sum_{i=1}^3[\Delta_i\hat S_{zi}+B_{\text{ac}}(\cos(\Delta_it)\hat S_{xi}+
\sin(\Delta_it)\hat S_{yi})]$.\\
Again, by means of the unitary transformation
$\hat U(t)=\text{exp}\{-i\sum_i\Delta_i\hat S_{zi}t\}$,
the states (\ref{bacstate})
turn out to still be eigenstates of the transformed $\hat
H'_{\text{B,2}}$, even if the two spins are now rotating in
resonance.
The bichromatic $B_{\text{ac}}$ washes out all the states with
antiparallel spins, driving the system into SB in spite
of the Zeeman inhomogeneity, see Fig.~\ref{fig5}.

\begin{figure}[t]
\begin{center}
\includegraphics[width=3in,clip]{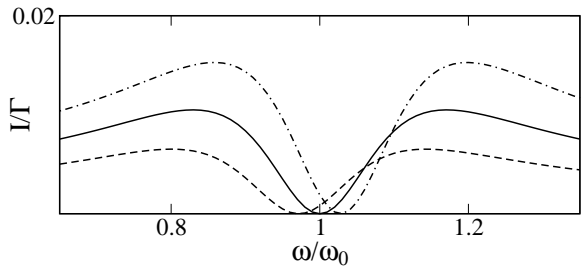}
\end{center}
\caption{\label{fig4}
\small $I$-$\omega$ in a TQD for $\Delta_1\ne\Delta_2$.
For $\tau \gg \Delta_1 - \Delta_2$, $I$ shows an antiresonance with a pronounced minimum at
$\hbar\omega_1 \approx \frac{1}{2}(\frac{\Delta_1 + \Delta_2}{2} + \Delta_3)$.
Dashed-dotted line: $\Delta_2 - \Delta_1 = 0.0013$; dashed line: $\Delta_1 - \Delta_2 = 0.0013$;
solid line: $\Delta_1 = \Delta_2$. Parameters: $\tau = \Gamma = 0.01$.}
\label{fig4}
\end{figure}

\begin{figure}[t]
\begin{center}
\includegraphics[width=3in,clip]{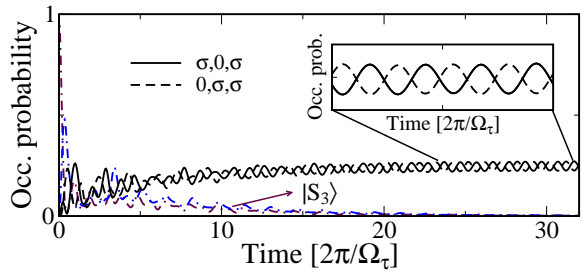}
\end{center}
\caption{\label{fig5}\small $\rho_{ii}$  in a TQD excited by
a bichromatic $B_{\text{ac}}$, where $\omega_1 = \Delta_{1,2}$ and
$\omega_2 = \Delta_3$. In the stationary limit, parallel spin states
decouple from $B_{\text{ac}}$ and form coherent superpositions as in
(\ref{bacstate})
thereby blocking the current.}
\end{figure}

{\it Conclusions} --- In summary, we have shown theoretically that TQD systems in triangular
configuration under dc and ac magnetic fields exhibit rich
dynamics due to the interplay of different coherent phenomena
induced by the magnetic fields.
For two extra electrons in the system
the interplay of Pauli exclusion principle and coherent trapping is
discussed in terms of the magnetic
flux piercing the TQD.
We have shown that, in contrast to the one electron case,
due to SB, electrons remain trapped even for $\Phi/\Phi_0 \ne n/2$.
We demonstrate that a generic property of
monochromatic and bichromatic magnetic fields is to induce SB 
at certain frequencies in both DQDs and TQDs.
Furthermore, the coherent superposition induced by the $B_{\text{ac}}$
constitutes a novel SB state and is decoupled from the field.
Its experimental realization will allow one to infer properties of the system
such as Zeeman inhomogeneities, and to manipulate spin qubits in DQDs and TQDs.
It opens new perspectives for manipulating spin transport properties,
thereby providing possibilities for designing spintronic devices.

We thank S. Kohler for critical reading of the manuscript
and T. Brandes, C. Emary and C.
P\"oltl for helpful discussions.
This work has been supported by grant
MAT2008-02626. M. B. acknowledges support from CSIC (JAE).
R.S. acknowledges support from Swiss NSF and MaNEP.

\end{document}